\documentclass[structabstract]{aa}

\usepackage{epsfig}
\usepackage{amssymb,amsmath}
\usepackage{txfonts}

\usepackage{natbib}
\bibpunct{(}{)}{;}{a}{}{,} 

\begin{document}

\title{Stellar scintillation in short exposure regime and atmospheric coherence time evaluation}
\author{Victor Kornilov}
\institute{Sternberg Astronomical Institute, Universitetskij pr-t, 13, 119992, Moscow, Russia }

\date{Received --- / Accepted ---}

\abstract {}{ Accurately measuring the atmospheric coherence time is still an important problem despite a variety of applicable methods. The Multi-aperture scintillation sensor (MASS) designed for the vertical profiling of optical turbulence, also provides a measurements of coherence time, but its results were found to be biased. Hence there is a need for a more robust method to determine $\tau_0$.}
{The effect of smoothing the stellar scintillation by a finite exposure of the detector is considered. The short exposure regime is described and its limits are defined. The re-analysis of previous measurements with the MASS is performed in order to test the applicability of this approach in real data processing. It is shown that most of the actual measurements satisfy the criteria of short exposures.}{ The expressions for the mean wind speeds $\bar V_2$ in the free atmosphere from the measurement of the scintillation indices  are derived for this regime. These values provide an estimate of the atmospheric coherence time $\tau_0$ without the need of empirical calibration. The verification of the method based on real measurements of the resulting $\tau_0$ are in good agreement with independent  methods.
}{}



\maketitle

\section{Introduction}
\label{sec:intro}

In recent decades the adequacy of \emph{optical turbulence} (OT) measurements in the Earth's atmosphere above potential astronomical sites and operating observatories has become more and more important. Developments in astronomy demand more in the way we determine OT. One such requirement is the ability to use cheap tools to monitor the OT on long time scales and automatically.

For such methods using a modest feeding optics, the analysis of the spatial wavefront distortion is replaced by the analysis of temporal variations of the wavefront inside a limited area. The widespread Differential Image Motion Monitor DIMM \citep{DIMM} which measures the turbulence integrated over the whole atmosphere, and the Multi-Aperture Scintillation Sensor MASS \citep{MASS,mnras2003} which determines the altitude distribution of the OT belong to such methods.

The transition from a spatial to a spatial-temporal description of the OT is based on the hypothesis of frozen turbulence \citep{Taylor} and was analyzed by many authors \citep[see e.g.][]{Roddier81,Martin1987}. A necessary component to this analysis is the knowledge of wind velocities in the atmosphere. 

The MASS instrument was developed to measure the vertical distribution of the OT. Its data contain the  necessary information about the temporal properties of stellar scintillation to estimate the \emph{atmospheric coherence time} $\tau_0$ with using a method involving the \emph{differential exposure scintillation index} DESI as described in \citet{timeconst2002}.

After the actual measurements of OT  with the MASS, it became clear that $\tau_0$ values derived in this way were significantly underestimated, and therefore required additional calibrations and corrections during analysis \citep{TMTVII}. Empirical recipes of correction  do not give confidence results because this systematic underestimation may be due to several factors \citep{TokoSite2010}. On the other hand, a large and permanently increasing volume of collected data requires an uniform treatment for objective comparison of OT characteristics  between various sites.

In this paper we analyse the effect of temporal averaging which affects on scintillation indices measured in the MASS instrument and their role in the evaluation of $\tau_0$. In addition the modification of the DESI method which does not depend on an empirical calibration is proposed. The presentation of the modified method is preceded by theoretical description of scintillation in the regime of short exposures.

The regime of short exposures allows us to establish a simple relationship between the measured scintillation statistics and the characteristic wind speed entering the definition of $\tau_0$. In the last section, the method is tested with MASS/DIMM data taken at Mt.~Shadzhatmaz \citep{kgo2010} and Mt.~Maidanak \citep{maid2005}.

We found that the main cause of the mentioned underestimations of $\tau_0$ was the wrong interpretation of the formula for DESI and these discrepancies virtually disappear after the correction in the data processing software.

The modified method does not require any empirical calibration. It is theoretically  clearer, and gives an estimate of the mean wind in the free atmosphere as well as a more accurate atmospheric coherence time.

\section{Temporal averaging of the scintillation}
\label{sec:temporal}

The theory of weak perturbations implies that the \emph{scintillation index} $s^2$ -- the variance of relative fluctuations of light intensity, is described by the sum of the scintillation indices produced by individual turbulent layers:
\begin{equation}
s^2 = \int_0^\infty C_n^2(h)\,W(h)\,{\rm d}h,
\label{eq:is}
\end{equation}
where $W(h)$ is the \emph{weighting function} (WF) which depends on the size and shape of the receiving aperture and does not depend on the altitude distribution of the structural coefficient of the refractive index $C_n^2(h)$. WF represents a power of the scintillation generated by a layer of unit intensity located at a height $h$.

The MASS method involves simultaneous measurements of the scintillation indices in 4 concentric apertures of different diameters (hereafter: A, B, C and D), leading to 10 independent scintillation indices. The vertical OT profile is restored from the measured indices and the theoretically calculated WFs \citep{mnras2003,MD2007}. The calculation of a set of functions $W(h)$ is based on the assumption that light intensity measurement has a \emph{zero exposure}, i.e. the averaging factor is related only to the receiving aperture.

The scintillation index $s^2$ measured with finite exposure time is determined by the integrated effect of all turbulent layers along the line of sight as in the case of zero exposure (\ref{eq:is}) but with another WFs.

The expression for the new WFs $W'(w,\tau,h)$ depends not only on the layer altitude $h$ but also on the wind speed $w = w(h)$ and averaging time $\tau$ \citep{timeconst2002,wind2010}.
It differs from the expression for $W(h)$ with an additional multiplicand  $A_s(w,\tau,f)$ in integrand:
\begin{equation}
W'(w,\tau,h) = 9.62\lambda^{-2}\int_0^\infty f^{-8/3} \sin^2(\pi r_F^2 f^2) \, A(f) \, A_s(w\tau f)\,{\rm d}f.
\label{eq:wdef}
\end{equation}
Here, $f$ is the modulus of the spatial frequency, $A(f)$ is spatial aperture filter (axisymmetric for MASS apertures), $r_F$ is the Fresnel radius $r_F^2 = \lambda h$.

Multiplicand $A_s(w \tau f) = A_s(\xi)$ is the spectral filter averaging with the wind shear $w\tau$  and can be expressed through the Bessel functions $J_0$ and $J_1$ and Struve functions $H_0$ and $H_1$  \citep{wind2010}:
\begin{equation}
A_s(\xi) = 2 J_0(2\pi\xi)-\frac{J_1(2\pi\xi)}{\pi\xi} - \pi J_0(2\pi\xi)H_1(2\pi\xi) + \pi J_1(2\pi\xi)H_0(2\pi\xi).
\end{equation}
It has a simple asymptotic behavior for small values of argument $\xi = w\tau f$: $A_s(\xi) \approx 1 - \pi^2\xi^2/6$, as it follows from its series expansion in the neighborhood of 0. The approximation provides an accuracy better than 0.02 until $\pi\xi < 1$. When $\xi \to \infty$, the function $A_s(\xi) \approx 1/\pi\xi$, and starting from  $\xi \approx 1$, the relative difference is less than 0.04.

These asymptotes correspond to the two extreme cases: the short and the long exposure regimes. Regime of \emph{short exposures} (SE) considered earlier by \citet{timeconst2002} will be the subject of this paper. Regime of \emph{long exposures} (LE) was studied recently by \citet{wind2010} in the application of potential impact of stellar scintillation on the accuracy of photometric measurements.

The main feature of these regimes is that the weighting function $W'(w, \tau, h)$ can be represented as the product of a function depending on wind shear, and the function which is independent on winds. It allows to separate the wind effect and the geometry of light propagation.

\section{Short exposures}
\label{sec:short}

In the expression (\ref{eq:wdef}) the integrand is significantly different from 0 in the region of intersection of the aperture ($f\lesssim 1/D$) and Fresnel ($f \lesssim 1/r_F$) filters. Outside this domain, i.e. when $f > \min\{1/D,1/r_F\}$ the integrand tends rapidly to 0. If the wind shear $\tau w \ll \max\{D,r_F\}$ then $f\tau w \lesssim 1$ and hence a quadratic approximation  $1 - \pi^2(w\tau f)^2/6$ of the spatial filter $A_s(w,\tau,f)$ can be applied.

\begin{figure}
\centering
\psfig{figure=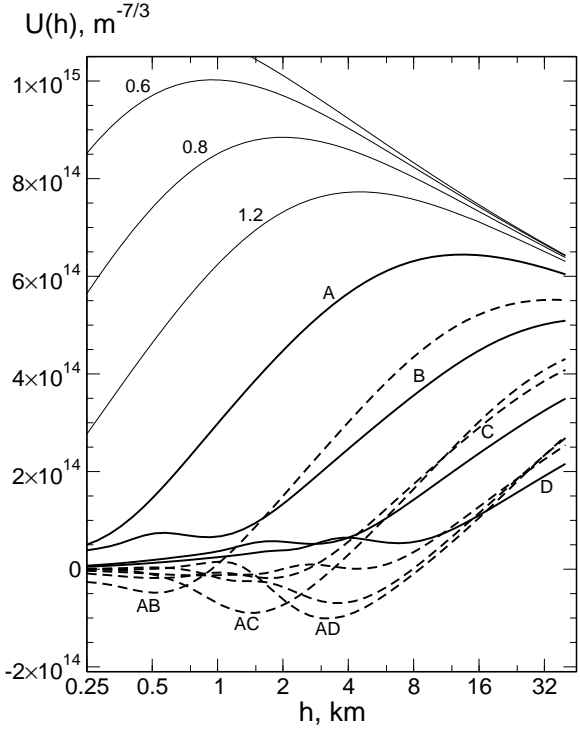,height=10cm}
\caption{Weighting functions $U(h)$ for a set of MASS/DIMM apertures computed for a typical spectral response of its detectors and the light source of spectral class A0\,V. Solid curves -- WFs for normal indices, dashed -- for cross-indices. Thin lines denote the functions for a number of smaller apertures, their diameters are marked in centimeters \label{fig:uf}}
\end{figure}

Let's replace $A_s$ for a certain aperture with the quadratic approximation in the expression (\ref{eq:wdef}) and take $w$ and $\tau$ outside the integral over frequency. Then
\begin{multline}
s^2_{\tau} = \int C_n^2(h)\,W'(w,\tau, h)\,{\rm d}h = \\ = \int C_n^2(h)\,W(h)\,{\rm d}h - \frac{\tau^2}{6}\int C_n^2(h)\,w(h)^2\,U(h) \,{\rm d}h,
\label{eq:sigt}
\end{multline}
In this difference the first integral is the scintillation power $s^2_0$ for zero exposure. The second integral (we denote it as $\mathcal V_2^U$) is the atmospheric second moment of wind additionally weighted with $U(h)$. Weighting function $U(h)$ is obtained by multiplying the initial spatial spectrum by $\pi^2 f^2$ and subsequently integrating over $f$. Since the effect of high-frequency spectral components increases after such multiplication the functions $U(h)$ become essentially different from the usually used $W(h)$ and have dimensions of $\mbox{m}^{-7/3}$. The set of $U(h)$ functions is shown in Fig.~\ref{fig:uf}.

Additional features of these WFs  are as follows:
\begin{itemize}
\item The spectral band width effect is more important than before owing to the high-frequency spectral components which are more intense. The difference of WFs for white and red stars reaches 10 \% in aperture A.
\item For infinitely small aperture an asymptote does not exist since at $D = 0$ the integral diverges. However, there is an envelope of $\sim h^{-1/6}$ to which the curves converge as $D\to 0$ (see Fig.~\ref{fig:uf}). 
\item The asymptote for the aperture $D \gg r_F$ is also very interesting: $U(h) \approx 17.22\,\lambda^{-2/3} \,D^{-3}\,h^{4/3}$. If compared to the asymptotic behaviour of the normal scintillation index, it is chromatic and has a stronger dependence on $D$. 
\end{itemize}

Note, that the WFs shown in the figure by the dashed lines are calculated for cross-indices, i.e. the values obtained directly from the MASS instrument measurements rather than differential indices. Differential indices were always calculated from the normal indices and covariances in the form of linear combinations, but from the viewpoint of the OT restoration it is a superfluous intermediate step which we have dismissed.

The SE regime  and the LE regime, are self-reproducible in the sense that if they occur in each turbulent layer they will be observed for the whole atmosphere. To determine the limits of SE regime, one can use the condition of applicability of the quadratic approximation of the wind shear filter $A_s(\xi)$, namely that at its border the function is equal to $1 - \pi^2\xi^2/6 = 5/6$.

Since for small $w\tau$ this function varies slowly with frequency $f$ within the main spectral peak (its maximum is located at frequency $\lesssim 0.7/r_F$) of the integrand, then the value of $s^2_{\tau}$ will also slightly vary with $w\tau$. Consequently we can assume that the quadratic approximation of the form (\ref{eq:sigt}) for a given aperture and atmospheric conditions is applicable while $\tau^2\,\mathcal V_2^U < s^2_0$ or
\begin{equation}
s^2_{\tau} > \frac{5}{6}s^2_0
\label{eq:condition2}
\end{equation}

Of course this inequality defines the SE regime only within some level of accuracy. One should take into account the error of approximation of the function $A_s(\xi)$ ($\approx 0.02$) and the contribution of the secondary peaks of the scintillation spectrum. A rough estimate of the accuracy of the numerical factor in (\ref{eq:condition2}) is $\approx 0.05$.

\begin{figure}
\centering
\psfig{figure=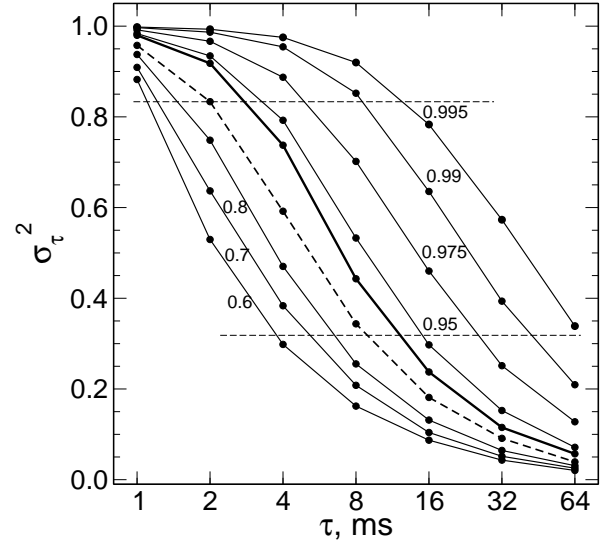,height=8.5cm}
\caption{Indices $s^2_{\tau}$ normalized to $s^2_0$ as function of an exposure $\tau$. Bold curve corresponds to the median of $\gamma_{12} = 0.938$, dashed line -- to the limit of the SE regime $\gamma_{12} = 0.870$. Another curves are marked by appropriate $\gamma_{12}$. The upper horizontal line represents the lower boundary of SE regime, the lower horizontal line -- the upper limit of the LE regime
\label{fig:examples}}
\end{figure}

Similarly it is possible to write the limit of the LE regime in the form of $s^2_{\tau} <\frac{1}{\pi} s^2_0$. Both boundaries are indicated in Fig.~\ref{fig:examples} with dashed lines. The Figure demonstrates examples of typical behavior of the measured indices $s^2_{\tau}$ with exposures ranging from 1 to 64~ms in  different wind conditions. Note that the left and right branch of each curve can be generated by different turbulent layers because in SE regime the contribution of an individual layer is $\sim C_n^2\,w^2$ while in the LE regime $\sim C_n^2/w$.

\section{Verification of applicability of SE regime in the MASS measurements}
\label{sec:test}

Since the $s^2_0$ value is calculated and not measured the restriction (\ref{eq:condition2}) is not directly applicable. However instead of $s^2_0$ one can use an additional index measured with a shorter exposure $\tau^\prime < \tau$. Simple algebraic manipulations lead to the condition
\begin{equation}
s^2_{\tau} > s^2_{\tau^\prime}\frac{5}{6-(\tau^\prime/\tau)^2}.
\label{eq:cond12}
\end{equation}
It follows that for exposures $\tau = 2$~ms and $\tau^\prime = 1$~ms $s^2_2$ should be more than $0.870\,s^2_1$, for 3 and 1~ms $s^2_3 > 0.850\,s^2_1$ and for 4 and 1~ms $s^2_4 > 0.842\,s^2_1$. Condition (\ref{eq:cond12}) can not be used when exposure $\tau^\prime$ itself is too long for the SE regime.

The analysis of actual MASS/DIMM data obtained at Mt.~Shatdzhatmaz in the period May 2009 to June 2010 was performed with the usual exposure $\tau = 1$~ms. The special version of MASS software recorded additional data in output files to calculate later the indices with exposures 2, 4, 8, 16, 32 and 64~ms (see Fig.~\ref{fig:examples}).

\begin{figure}
\centering
\psfig{figure=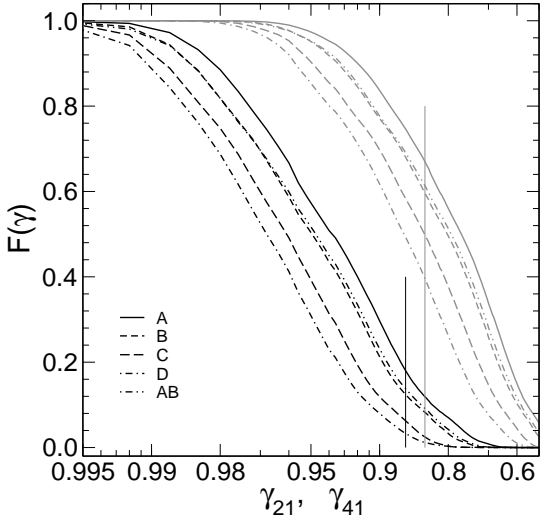,height=8cm}
\caption{Cumulative distributions of $\gamma_{21}$ (black curves) and $\gamma_{41}$ (grey curves) for the apertures A, B, C, D and AB cross-index based on measurements at Mt.~Shatdzhatmaz.
Vertical lines cut off points which do not satisfy the SE regime
\label{fig:gamma}}
\end{figure}

Cumulative distributions of ratios $\gamma_{21} = s^2_2/s^2_1$ and $\gamma_{41} = s^2_4/s^2_1$ averaged over a minute are shown in Fig.~\ref{fig:gamma}. As it might be expected from the form of the WFs (Fig.~\ref{fig:uf}) smaller ratios are observed for smaller apertures, more sensitive to OT motion, and the distribution for AB cross-index is very close to the curve for the aperture B. Numerical characteristics of the distributions are given in Table~\ref{tab:proc} and show that the 2-ms measurement is almost always within the SE regime. Only 18\% of the measurements in A aperture fail to meet the $\gamma_{21} > 0.870$ condition.

Characteristics of the distributions can vary greatly from season to season. E.g. during February--March 2010 period, which was characterized by strong winds, the median $\gamma_{21}$ decreased to 0.90, while the fraction of measurements that do not satisfy the criterion of SE increased to 38\%. On the other hand, all measurements fell into SE regime in October 2009 which was notable for its stable calm weather.

\begin{table}
\small
\caption{Characteristics of $\gamma_{21}$ and $\gamma_{41}$ distributions for measurements at Mt.~Shadzhatmaz in May 2009 -- June 2010 period\label{tab:proc}}
\bigskip
\centering
\begin{tabular}{lrrrrr}
\hline\hline
Parmeter \rule{0pt}{12pt}     & A & B & C & D & AB \\[2pt]
\hline
\rule{0pt}{12pt}    & \multicolumn{5}{c}{Distribution of $\gamma_{21}$} \\[2pt]
Median              & 0.94 & 0.95 & 0.96 & 0.97 & 0.95 \\
Out of SE regime,\%    & 18.0  & 12.5  & 6.3   & 3.3   & 13.8 \\[2pt]
                    & \multicolumn{5}{c}{Distribution of $\gamma_{41}$} \\[2pt]
Median              & 0.77 & 0.80 & 0.84 & 0.87 & 0.79\\
Out of SE regime,\%    &  67.0 &  59.6 &  49.8 &  39.6 &  61.3\\[2pt]
\hline
\end{tabular}
\end{table}

\begin{table}
\small
\caption{Characteristics of $\gamma_{21}$ and $\gamma_{31}$ distributions for measurements at Mt.~ Maidanak in August 2005 -- December 2007 period\label{tab:procm}}
\bigskip
\centering
\begin{tabular}{lrrrr}
\hline\hline
Parmeter \rule{0pt}{12pt}   & A & B & C & D  \\[2pt]
\hline
\rule{0pt}{12pt}    & \multicolumn{4}{c}{Distribution of $\gamma_{21}$}\\[2pt]
Median              & 0.92 & 0.93 & 0.94 & 0.96  \\
Out of SE regime,\%    & 19.4  & 13.3  & 5.5   & 0.3  \\[2pt]
                    & \multicolumn{4}{c}{Distribution of $\gamma_{31}$}\\[2pt]
Median              & 0.82 & 0.84 & 0.86 & 0.91  \\
Out of SE regime,\%    & 64.1 &  57.6 &  41.5 &  12.8 \\[2pt]
\hline
\end{tabular}
\end{table}

For measurements with 4~ms exposure (value $\gamma_{41}$) the situation is radically different. In this case, the vast number of measurements does not satisfy the SE condition and consequently the quadratic approximation can be used only in rare cases and with caution.

Similar analysis was performed for measurements at Mt.~Maidanak. The standard MASS data were used, from which indices $s^2_2 = (s^2_1 + \rho_1)/2$ and $s^2_3 = (3s^2_1 + 4\rho_1 + 2\rho_2)/9$ were calculated to construct cumulative distribution of $\gamma_{21}$ and $\gamma_{31} = s^2_3/s^2_1$. General properties of these distributions do not differ from that of Mt.~Shatdzhatmaz. Some differences are due to significantly larger apertures C and D of the first generation device \citep{MASS}. Characteristics of the distributions are given in Table~\ref {tab:procm}. The fraction of measurements dropping out of SE regime in the case of A aperture is 19.4\% for exposure 2-ms and 64\% for 3-ms exposure.

In the situation where $\gamma_{21}$ approaches unity, a very high precision of the ratio is required as the curves for $\gamma_{21} = 0.995$ and $\gamma_{21} = 0.99$ differ very much. For these curves even 8~ms measurements are in the SE regime. 

\section{Reduction to zero exposure}
\label{sec:corr0}

Typical exposures of 1~ms are taken with the MASS. This leads to a wind shear of the order of 3~cm assuming wind speed in the tropopause $\sim 30\mbox{ m s}^{-1}$.  The value is comparable with the size of the device apertures and with typical Fresnel radius so 1~ms exposure can not be considered as infinitely small one. The procedure used to correct indices to zero exposure is provided during the MASS data processing. The algorithm is based on numerical simulations \citep{timeconst2002}.

However for the measurements in SE regime, the required correction can be calculated with the help of a direct method using the two indices $s^2_1$ and $s^2_2$ obtained with different exposures. Although the ratio of exposures may be arbitrary, 
it is convenient to consider the case of a single $\tau$ and a double $2\tau$ exposures. Then the expression (\ref{eq:sigt}) becomes:
\begin{equation}
s^2_1 = s^2_0 - \frac{\tau^2}{6}\mathcal V_2^U, \qquad s^2_2 = s^2_0 - \frac{4\tau^2}{6}\mathcal V_2^U.
\label{eq:sig2}
\end{equation}
After solving this system for $s^2_0$ we obtain
\begin{equation}
s^2_0 = \frac{4}{3}s^2_1 - \frac{1}{3}s^2_2.
\label{eq:2tau}
\end{equation}
One can express $s^2_2$ in terms of $s^2_1$ and covariance $\rho_1$ of adjacent counts as $2s^2_2 = s^2_1 + \rho_1$. The formula can be re-written as:
\begin{equation}
s^2_0 = \frac{7}{6}s^2_1 - \frac{1}{6}\rho_1.
\label{eq:2rho}
\end{equation}
The resulting correction is somewhat smaller than the one adopted in \citet{MD2007}: $s^2_0 = 1.25s^2_1 - 0.25\rho_1$. Of course, formulas (\ref{eq:2tau}) and (\ref{eq:2rho}) are exact only if the condition (\ref{eq:condition2}) of SE regime is satisfied for $s^2_2$.

The statistical error of $\rho_1$  is very close to that of $s^2_1$. Consequently the correction of the scintillation index increases its standard error by a factor of 1.18 which is not significant in practice. 
%

\section{Atmospheric second moment of wind}
\label{sec:wind2}

We can estimate the integral $\mathcal V_2^U$ from the system of equations (\ref{eq:sig2}). Solving the system for the indices measured in the $j$-aperture ($j = 1, \dots, 10$) with exposures $\tau_1$ and $\tau_2$ relative to this unknown, we obtain:
\begin{equation}
\mathcal V_2^{U_j} = \int C_n^2(h)\,w(h)^2\,U_j(h) \,{\rm d}h = 6\,\frac{s^2_{\tau_1} - s^2_{\tau_2}}{\tau_2^2-\tau_1^2} = \Delta_j,
\label{eq:apj}
\end{equation}
where $\Delta_j$ denotes the measured quantity. The integral is of little interest by itself because it includes an additional factor $U_j(h)$ which distorts the contribution of different altitudinal layers. Real estimation of the second atmospheric moment of the wind $\mathcal V_2 = \int C_n^2(h)\,w(h)^2\,{\rm d}h$ can be obtained if we find a linear combination $A_U(h)$ of functions $U_j(h)$ which is close to unity.

\begin{figure}
\centering
\psfig{figure=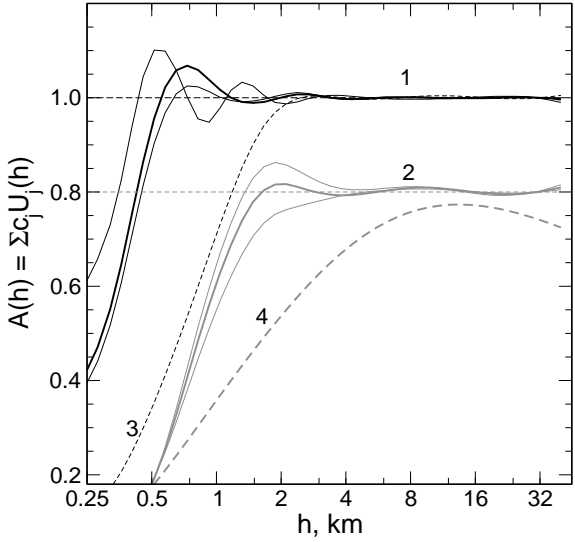,height=8.5cm}
\caption{Approximation $A_U(h)$ of a constant level with help of the set of $U(h)$. Black curves (1) show the approximation with 10 indices, grey (2) -- with 4 indices. For clarity the grey graphs are shifted to the level of 0.8. Thick lines correspond to the weight of $\sim h$, thin -- to other power of altitude. Curve 3 corresponds to the weighting of $\sim h^{7/5}$. Gray dashed 4 -- WF for the aperture A
\label{fig:app-uf}}
\end{figure}

Such combination is shown in Fig.~\ref{fig:app-uf}. Good results can even be obtained when only using functions corresponding to normal indices. The method and results of the decomposition $A_U(h) \approx 1$ are described in detail in Appendix~A.

Expansion coefficients of $A_U(h) =\sum_j c_j U_j(h)$ are of order of $10^{-15}\mbox{ m}^{7/3}$. Summing equation (\ref{eq:apj}) for the indices with these coefficients we obtain:
\begin{equation}
\sum_j c_j \Delta_j \approx \mathcal V_2.
\label{eq:sum}
\end{equation}

It is clear that the boundary layer ($h < 0.5$ km) wind will not be accounted for fully in the integral and the surface wind will be completely excluded despite the strong surface turbulence. The behavior of  $A_U(h)$ is similar to the approximating function under the integral defining the intensity of turbulence in the free atmosphere $J_{free}$. Therefore this quantity can be used for further normalization. The contribution of the surface layer can be accounted for later on base of DIMM data.

Dividing both sides of the expression (\ref{eq:sum}) by $J_{free}$, we obtain the mean square of the wind speed $\langle w^2 \rangle$ in the free atmosphere and its expression in terms of measurable quantities:
\begin{equation}
\langle w^2 \rangle =\frac{\mathcal V_2}{J_{free}} = \frac{\sum_j c_j \Delta_j}{J_{free}}.
\end{equation}

\section{Atmospheric coherence time}
\label{sec:tau0}

It was shown by \citet{TKel2007} that the mean wind $\bar V_2$ can be a good estimator of $\bar V_{5/3}$ which is used in the definition of the atmospheric coherence time $\tau_0$. Recalling the fact that the instrument FADE \citep[Fast Defocusing of stellar image,][]{FADE2008} measures the time constant (interferometric coherence time) also using $\bar V_2$. Although authors indicate that on average $\bar V_2 \approx 1.1\bar V_{5/3}$, we will assume that both values are equivalent in the following argumentation.

Evaluation of $\bar V_2$ (in the free atmosphere) from the measurements is obtained directly from the formula for $\langle w^2 \rangle$:
\begin{equation}
\bar V_2 = \langle w^2 \rangle^{1/2} = \frac{(\sum_j c_j \Delta_j)^{1/2}}{J_{free}^{1/2}}.
\end{equation}
Substituting this value in the definition of the atmospheric time constant $\tau_0$ we obtain the following expression:
\begin{equation}
\tau_0 = 0.314\frac{r_0}{\bar V_2} = 0.058\lambda^{6/5} J_{tot}^{-3/5}\frac{J_{free}^{1/2}}{(\sum_j c_j \Delta_j)^{1/2}},
\end{equation}
where we have substituted the Fried parameter $r_0$ with OT intensity $J_{tot}$ in the whole atmosphere. For the wavelength $\lambda = 500$~nm:
\begin{equation}
\tau_0 = 1.593\cdot10^{-9} J_{tot}^{-3/5}\frac{J_{free}^{1/2}}{(\sum_j c_j \Delta_j)^{1/2}}.
\label{eq:tau0eval}
\end{equation}

If there is information about the OT power in the boundary layer and surface wind speed $V_0$, then its contribution can be accounted for at the stage of calculating the mean wind $\bar V_2$. In order to do this, we correct the mean square of the wind speed by adding $V_0^2$ in proportion of the surface layer intensity $J_{GL}$:
\begin{equation}
\tilde{\langle w^2 \rangle} = \frac{\langle w^2 \rangle J_{free} + V_0^2 J_{GL}}{ J_{tot}},
\end{equation}
or
\begin{equation}
\bar V_2 = \frac{\left(\sum_j c_j \Delta_j + V_0^2 J_{GL}\right)^{1/2}}{J_{tot}^{1/2}}.
\end{equation}
Substituting the corrected value in the formula (\ref{eq:tau0eval}) we obtain an estimation of the atmospheric coherence time for the whole atmosphere:
\begin{equation}
\tau_0 = 1.593\cdot10^{-9} \frac{J_{tot}^{-1/10}}{(\sum_j c_j \Delta_j + V_0^2 J_{GL})^{1/2}}.
\label{eq:tau0whole}
\end{equation}
Note that the usual method of adding coherence time $\tau_0^{-5/3} = \tau_{GL}^{-5/3} + \tau_{free}^{-5/3}$ is not correct if used to calculate $\bar V_2$ which has square metric.

The relative accuracy of $\tau_0$ does not practically depend on the accuracy of the $J_{tot}$, and is twice as good as the accuracy of the measured $\Delta_j$. 

\section{Verification of the method of evaluation $\tau_0$}
\label{sec:verific}

The method presented here was tested with data  obtained at Mt.~Shadzhatmaz and Mt.~Maidanak which were analyzed in Sect.~\ref{sec:test}. To construct $\sum_j c_j \Delta_j$ we have used the coefficients from the Table~\ref{tab:decomp} defined for stars of spectral class A0\,V. This simplification introduces a systematic error when processing the measurements of stars another spectral classes so further results should not be regarded as final (less than 5\%, see Sect.~\ref{sec:short}). We also computed the coherence time for free atmosphere, i.e. $J_{free}$ value was used instead of $J_{tot}$ in formula (\ref{eq:tau0eval}). 

\begin{figure}
\centering
\psfig{figure=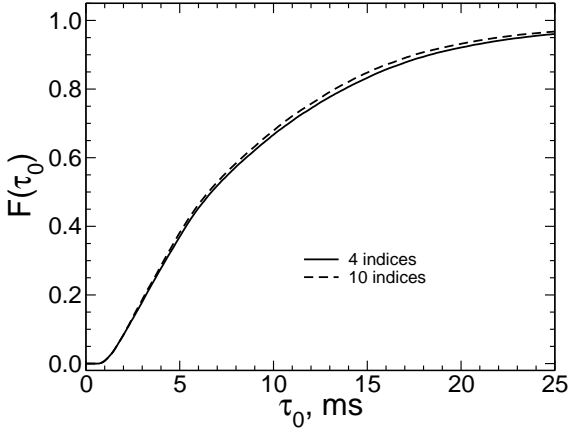,height=7.8cm,angle=-90}
\caption{Cumulative distribution of free atmosphere $\tau_0$ from measurements at Mt.~Shatdzhatmaz obtained using $s^2_1-s^2_2$ differences with decomposition in 4 (solid line) and 10 (dashed line) indices 
\label{fig:tau0}}
\end{figure}

For the measurements at Mt.~Shadzhatmaz differences $\Delta_j$ were computed from indices $s^2_1$ and $s^2_2$. The results of the evaluation are shown in Fig.~\ref{fig:tau0} where the cumulative distributions of $\tau_0$ obtained with different decompositions are presented.
The method results in 6.69~ms median when using 4 normal indices and 6.53~ms using all 10 indices. The difference between these distributions is completely explained by the fact that usage of all the indices involves additional contribution of moving turbulence between 0.5 and 1~km.


\begin{figure}
\centering
\psfig{figure=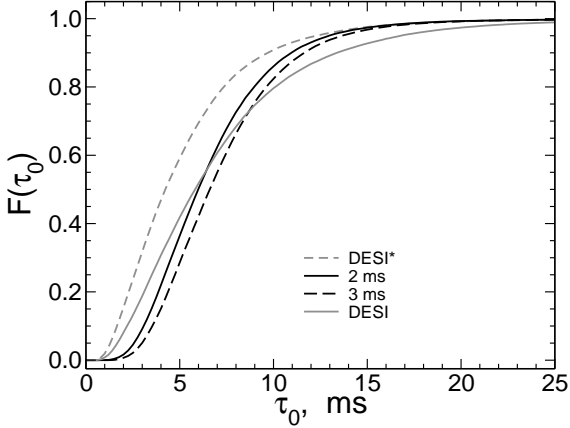,height=7.8cm,angle=-90}
\caption{Cumulative distribution of free atmosphere $\tau_0$ from measurements at Mt.~Maidanak obtained using DESI (grey dashed -- before software correction, grey solid -- after) and using $s^2_1-s^2_2$ and $s^2_1-s^2_3$  differences (black solid and dashed curves respectively) \label{fig:maid_tau}}
\end{figure}

In the case of the standard MASS output (Maidanak campaign) normal indices $s^2_1$, $s^2_2$ and $s^2_3$ were used to study biasing of $\tau_0$ estimations with different temporal span. At first we investigated the effect of  the DESI formula correction, i.e. replacement of the index $s^2_0$ with the index $s^2_1$. 
As can be seen from Fig.~\ref{fig:maid_tau} the correction increases $\tau_0$ estimations by a factor $\approx 1.4$. The cumulative distributions of coherence time calculated from $s^2_1 - s^2_2$ and $s^2_1 - s^2_3$ are presented in the Figure too. One can see that for the case of corrected DESI algorithm (5.82~ms) and 2~ms median (5.98~ms), the median differs only slightly.

The $\tau_0$ values obtained with $s^2_3$ are systematically larger by $\approx 0.6$~ms than $\tau_0$ calculated with $s^2_2$. This shift is caused by the 3~ms exposure beyond the SE regime. The larger $\tau_0$ gets the more 3~ms measurements satisfy the criterion of SE and because of that, the relative difference between the curves for 2~ms and 3~ms is reduced, becoming less than 4\% at $\tau_0 \approx 15$~ms. To obtain the systematic shift of the 2~ms curve, additional measurements with a shorter exposure are required. However taking into account the quadratic dependence of indices on exposure, we can estimate that its median is not overstated by more than 5\%.

Random errors of $\tau_0$ were computed during the averaging of $\bar V_2$ over 1 minute as the standard error of the mean. The relative error of $\bar V_2$ is approximately constant in entire range of values. At very low winds (less than $5\mbox{ m s}^{-1}$) errors increase. The median of relative error is $\approx 0.02$, only 1\% of the measurements have errors greater than 0.1.

\section{Discussion and Conclusion}

The analysis of actual MASS measurements described in Sect.~\ref{sec:test}, has shown that for the commonly used 1 ms exposure 80\% of cases, are within the SE regime and that the correction to zero exposure (\ref{eq:2rho}) is suitable. Despite some conventionality of the regime boundary definition, it can not be significantly weakened. On the contrary the study of real data requires some tightening of the limit to guarantee that all turbulent layers are within the SE regime. Our statistics shows that exposures of 3~ms rarely satisfy the criterion of SE.

Such a control is important because measurements of $\tau_0$ in large wind shear will be affected by systematic errors. Fortunately such cases are not as interesting as opposite ones.

The reaction of the DESI method on fast turbulence motion is not so evident. One can derive from the definition of the differential exposure signal \citep{TokoSite2010} and formula (\ref{eq:apj}):
\begin{equation}
DESI = \frac{2}{9}(3s^2_1 - 4\rho_1 + \rho_2 ) = \frac{3}{4}(\Delta_A^{12} - \Delta_A^{13}).
\end{equation}
This means that the DESI method estimates the coherence time not by using the atmospheric wind moment but its derivative with exposure. DESI differences are smaller by an order of magnitude than the values themselves and therefore the DESI method is noisier. Indeed, for Mt.~Maidanak data the median of relative random errors is $\approx 0.03$ using the difference $s^2_1-s^2_2$ while using DESI the median of relative errors amounts $\approx 0.1$.

The case of calm atmosphere is most interesting from the viewpoint of adaptive optics and interferometry. In this situation indices differences are small and poorly defined owing to an increase of relative error. The most dangerous is a systematic error due to an incorrect accounting for photon noise. To reduce the impact of such errors, an adaptable choice of exposure may be realized \citep{TokoSite2010}.

Fortunately, in real situations this effect is not too large. For example measurements at Mt.~Maidanak show that error $\Delta p = 0.01$ in the parameter $p$ for photon noise correction changes $\tau_0$ median by 0.05~ms. At $\tau_0 \approx 15$~ms (95\% level of cumulative distribution) this error results in shift of the curve by 0.7~ms when used $s^2_1-s^2_2$ and 0.3~ms when $s^2_1-s^2_3$.

Generally speaking, the SE regime is not very suitable to determine wind conditions. Much more accurate results can be obtained by the determination of the averaging time $t$ for which the index $s_t^2$ becomes equal to $s_0^2/2$. Such method is similar to the calculation of correlation peak width, see e.g., \citet{caccia1987}. However, in the middle domain (see Fig.~\ref{fig:examples}) the behavior of $s_t^2$ indicates that similar integral characteristic are closer to mean wind speed $\langle w\rangle$ rather than to the mean square $\langle w^2 \rangle$. The mean speed may be used for some applications but for $\tau_0$ evaluation the value $\langle w^2 \rangle$ is more suitable.


The analyzed data were obtained at sites with rare high-altitude jet streams. For other observatories, especially closer to the equator, the situation could be worse. In such case, it makes sense to reduce the exposure to 0.5 -- 0.7~ms, which is equivalent to increasing the wind range 1.5 -- 2 times. The main objective of a correct choice of exposure is to ensure that we are within the SE regime for statistically overwhelming fraction of cases.

The main advantage of making measurements in the SE regime is the simplicity of theoretical description and its practical usage. The described method for $\tau_0$ evaluation does not contain any empirical calibration, it gives more accurate estimations and additional information about mean wind in the free atmosphere.

All quantities on the right hand side of (\ref{eq:tau0eval}) are measured with MASS/DIMM simultaneously. However the joint MASS/DIMM data processing was realized only in the recent version of the program \emph{atmos} \citep{kgo2010} and the algorithm described in Sect.~\ref{sec:tau0}, will be implemented in the next release of this version. An alternative way is to use a fairly simple post-processing of existing MASS output, but in both cases the meteo data on the surface wind is additionally needed to take the ground layer into account. Free atmosphere $\tau_0$ evaluation using both methods is implemented starting with \emph{atmos-2.97.3} version.

\begin{acknowledgements}
The author thanks his colleagues who took part in campaigns of optical turbulence measurements at Mt.~Maidanak and Mt.~Shatdzhatmaz, by whose efforts the data set used in the analysis was collected. Concerned discussion on the atmospheric coherence time from MASS data with participants of the conference ``Comprehensive characterization of astronomical sites'' also stimulated the implementation of this work. The author is particularly grateful to A.\,Tokovinin and N.\,Shatsky for useful discussions, T\,Travouillon for his efforts to make the paper clearer and M.Sarazin for his interest in the appearance of this work.
\end{acknowledgements}

\bibliographystyle{aa}
\bibliography{reference_list}

\appendix
\section{Decomposition in the weighting functions}
\label{app:app}

The decomposition of the atmospheric wind moment $\mathcal V^2$ is done in the same way as an expansion of altitude atmospheric moments with sets of measured indices \citep{mnras2003} in the program \emph{atmos}. In this case we solve the system of linear equations $ \tens{U} {\bf c} = {\bf 1}$, where $\tens{U}$ is the WFs matrix with dimension $k \times n$, ${\bf c}$ is the vector of coefficients, ${\bf 1}$ is the unit vector corresponding to the altitude grid ${\bf h}$. The number of nodes $n = 50$ is significantly greater than the maximal number of indices $k = 10$.  As usually log-uniform grid is used which has more nodes at low altitudes.

The system is solved by SVD with regularization of the solution by discarding small singular values $s_i < 10^{-3}\,s_0$. The quality of the solution was controlled by two parameters: maximum deviation from 1 (not accounting the initial part of the curve) and noise enhancement factor $NF = (\sum_j c_j^2)^{1/2}/\sum_j c_j$. The solution depends on how the system was weighted. We used an implicit weight by grid density and some explicit weight.

Without explicit weight, the solution has the first maximum at the minimum altitude but strongly oscillates and has large $NF > 20$. The study of additional weight in the form of $\sim h^p$ has shown that if $p \approx 1$ the solution is good enough: it is close to 1 and has moderate $NF$ value.
\begin{table}
\small
\caption{Coefficients $c_j$ ($10^{-15}\mbox{ m}^{7/3}$) of $h^0$ decomposition in WFs $U(h)$ when we used all indices, 4 normal indices, 4 normal indices for original MASS device ---columns 2, 3, 4 respectively \label{tab:decomp}}
\bigskip
\centering
\begin{tabular}{crrr}
\hline\hline
Aperture\rule{0pt}{11pt}& $c_j$  & $c_j$ & $c_j$ \\[2pt]
\hline
A \rule{0pt}{11pt}      & 3.229     &  2.981  & 2.504    \\
B                       & 3.191     & $-3.641$& $-2.823$  \\
C                       & 0.080     & 2.880   & 2.960    \\
D                       &$-0.826$   & 0.273   & $-0.730$  \\
AB                      &$-5.040$   & $\dots$ & $\dots$   \\
AC                      & 1.124     & $\dots$ & $\dots$   \\
AD                      & 0.013     & $\dots$ & $\dots$   \\
BC                      &$-0.810$   & $\dots$ & $\dots$   \\
BD                      & 0.547     & $\dots$ & $\dots$   \\
CD                      & 0.166     & $\dots$ & $\dots$   \\[2pt]
\hline
Noise Factor\rule{0pt}{11pt} & 4.16 & 2.21 & 2.54  \\[2pt]
\hline
\end{tabular}
\end{table}

The approximating curves $A_U(h)$ calculated with all 10 indices and only 4 normal indices are shown in Fig.~\ref{fig:app-uf}. The curve for aperture A  for which an index and temporal covariances are used in the DESI method to assess the atmospheric coherence time is also shown. The coefficients $c_j$ are presented in Table~\ref{tab:decomp} for all variants. They were calculated for a typical set of MASS/DIMM apertures, its detector response and A0\,V spectral class of light  source.

The approximation which uses all indices is close to 1 for all nodes of the typical grid for restoration of OT profile. There is a slight ($\approx 0.06$) excess at the altitude of 0.7~km and the difference for other nodes is lower. The approximation with only 4 normal index looses 25\% of the contribution  of 1~km layer and almost 75\% of the contribution of 0.5~km layer. It is expected that the decomposition in the complete set of indices will give an estimation of $\tau_0$ with smaller systematic errors because the noise properties of cross-indices are not worse than that of normal indices and the factor $NF$ in the first variant is only twice as large.

\end{document}